# Deficiency of the Bulk Spin Hall Effect Model for Spin-Orbit Torques in Magnetic Insulator/Heavy Metal Heterostructures

Junxue Li[1#], Guoqiang Yu[2#], Chi Tang[1], Yizhou Liu[3], Zhong Shi[1,4], Yawen Liu[1], Aryan Navabi[2], Mohammed Aldosary[1], Qiming Shao[2], Kang L. Wang[2], Roger Lake[3] and Jing Shi[1*]

1. Department of Physics and Astronomy, University of California, Riverside, California, USA
2. Department of Electrical Engineering, University of California, Los Angeles, California, USA
3. Department of Electrical and Computer Engineering, University of California, Riverside, California, USA
4. School of Physics Science and Engineering, Tongji University, Shanghai, China

#These authors contributed to this work equally.

Electrical currents in a magnetic insulator/heavy metal heterostructure can induce two simultaneous effects, namely, spin Hall magnetoresistance (SMR) on the heavy metal side and spin-orbit torques (SOTs) on the magnetic insulator side. Within the framework of the pure spin current model based on the bulk spin Hall effect (SHE), the ratio of the spin Hall-induced anomalous Hall effect (SH-AHE) to SMR should be equal to the ratio of the field-like torque (FLT) to damping-like torque (DLT). We perform a quantitative study of SMR, SH-AHE, and SOTs in a series of thulium iron garnet/platinum or $Tm_3Fe_5O_{12}$/Pt heterostructures with different $Tm_3Fe_5O_{12}$ thicknesses, where $Tm_3Fe_5O_{12}$ is a ferrimagnetic insulator with perpendicular magnetic anisotropy. We find the ratio between measured effective fields of FLT and DLT is at least 2 times larger than the ratio of the SH-AHE to SMR. In addition, the bulk SHE model grossly underestimates the spin torque efficiency of FLT. Our results reveal deficiencies of the bulk SHE model and also address the importance of interfacial effects such as the Rashba and magnetic proximity effects in magnetic insulator/heavy metal heterostructures.

Manipulation of magnetization with pure spin current in ferromagnet (FM)/heavy metal (HM) heterostructures has attracted a great deal of attention from both fundamental and application perspectives. It has been experimentally established that a charge current in the HM film with strong spin-orbit coupling can produce sufficiently strong spin-orbit torques (SOTs) to cause precession [1-3] or even switching [4-10] of the FM magnetization of an adjacent layer. Phenomenologically, two types of SOTs are proposed to describe the experimental results, i.e., damping-like torque (DLT) directed along $\hat{m} \times (\hat{m} \times \hat{\sigma})$ and field-like torque (FLT) along $\hat{m} \times \hat{\sigma}$, where $\hat{m}$ and $\hat{\sigma}$ represent the unit vectors of magnetization $M$ and spin polarization, respectively. Fundamentally, both bulk and interface effects can give rise to these torques, the former from the spin Hall effect (SHE) [11] in the HM layer and the latter from the Rashba [12, 13] and other effects at the FM/HM interface [14-16]. Despite extensive theoretical and experimental studies so far [5-10], a consensus is still lacking as to which mechanism is mainly responsible.

In magnetic insulator (MI)/HM heterostructures, the situation is very different. Conduction electrons in the HM layer do not enter the MI layer; therefore, the s-d exchange and consequently spin angular momentum transfer occurs right at the interface. The spin current generated by SHE in the HM layer transmits through the interface which is quantitatively described by the spin-mixing conductance [17], and as a result produces the same form of DLT and FLT as in the metallic FM/HM heterostructures. In the meantime, the spin current transmission through the interface is connected to the transport phenomena in the HM layer such as the spin Hall-induced anomalous Hall effect (SH-AHE) and spin-Hall magnetoresistance (SMR) [18-21]. Therefore, the same complex spin-mixing conductance links both SOTs and SH-AHE/SMR, if only the bulk spin current is present [18, 19, 22]. The bulk SHE theory leads to the following relations between the effective fields, i.e. $H_{FL}$ and $H_{DL}$ for the corresponding FLT and DLT, and the magnitude of SH-AHE and SMR, i.e., $H_{FL} \propto \frac{\Delta \rho_{AHE}^{sh}}{\rho}$ and $H_{DL} \propto \frac{\Delta \rho_{SMR}}{\rho}$ (see Supplement Material, or SM [23]), where $\rho$ is the electrical resistivity of the HM layer, $\Delta \rho_{AHE}^{sh}$ and $\Delta \rho_{SMR}$ are the resistivity modulations of the SH-AHE and SMR, respectively. Consequently, we have

$$\frac{\Delta \rho_{AHE}^{sh}}{\Delta \rho_{SMR}} = \frac{H_{FL}}{H_{DL}}. \qquad (1)$$

This equation is expected to hold only if interfacial effects such as the Rashba spin-orbit coupling and magnetic proximity effect [15, 16] are negligible. In practice, the interfacial effects are not only ubiquitous, but may also play an important role in MI/HM heterostructures, which likely leads to the breakdown of Eq. 1. Hence, this relation serves as the first validity test of the bulk SHE-based model but has not been experimentally carried out yet. A major challenge is the lack of direct electrical response from the MI layer itself. Montazeri et al. [24] reported a magneto-optical investigation of SOTs in yttrium iron garnet ($Y_3Fe_5O_{12}$ or YIG)/Pt bilayer structures and extracted a relatively large DLT but a negligible FLT. Recently, Li et al. [25] and Avci et al. [26] reported DLT-induced switching in MI with perpendicular magnetic anisotropy (PMA) by taking advantage of the SH-AHE signal in MI/HM bilayers. However, quantifying the magnitude of much smaller FLT in MI/HM remains challenging.

In this Letter, we focus on the determination of both DLT and FLT and the correlation between SH-AHE/SMR and SOTs in thulium iron garnet ($Tm_3Fe_5O_{12}$ or TIG)/Pt heterostructures. By performing magneto-transport and angle-dependent harmonic Hall measurements, we extract the magnitude of the anomalous Hall resistivity, SMR, and the effective fields corresponding to both SOTs in the TIG/Pt heterostructures with varying TIG thickness. Within the framework of the bulk SHE, we attribute the anomalous Hall resistivity solely to the SH-AHE. We find that $H_{FL}/H_{DL}$ is at least 2 times larger than $\Delta\rho_{AHE}^{sh}/\Delta\rho_{SMR}$, and the spin torque efficiency for FLT determined from the SOT measurements is much larger than the estimated value from the SH-AHE. These large discrepancies suggest that the bulk SHE model grossly underestimates FLT and interfacial effects must be considered.

We grow TIG films on single crystalline $Nd_3Ga_5O_{12}$ (111) substrates using pulsed laser deposition. Growth details can be found in our previous work [16]. The lattice-mismatch induced tensile strain and negative magneto-crystalline anisotropy coefficient [27] in TIG drive its magnetization perpendicular to the film plane, which is directly characterized by vibrating sample magnetometry and further confirmed by the squared out-of-plane anomalous Hall hysteresis loops in TIG/Pt. In this work, TIG films with thickness of 3.2, 4.8, 6.4 and 9.4 nm are prepared followed by the deposition of 4-nm-thick Pt films by sputtering (the stack structure is shown in the inset of Fig. 1(b)). For transport measurements, the TIG/Pt heterostructures are patterned into Hall bars (as depicted in Fig. 1(a) and the inset of Fig. 1(b)) using standard photolithography and dry etching. We measure both longitudinal and transverse resistances using a DC current with an amplitude of 2.0 mA, and the harmonic Hall signals using an alternating current (AC) with a frequency of 13.117 Hz. All measurements are performed at room temperature. We have carried out magneto-transport measurements on all four samples, but here we primarily present the results from TIG(4.8 nm)/Pt(4 nm) heterostructure to demonstrate detailed analysis. It is worth mentioning that we have observed SOT-induced magnetization switching in all devices (see SM [23]), but in this Letter we only focus on testing the validity of the bulk SHE mechanism for SMR and SOTs.

First, we show magneto-transport measurements in TIG/Pt. In Fig. 1(b), the sharp squared Hall loop clearly indicates robust PMA of the TIG film. The magnitude of the anomalous Hall resistivity is $\Delta\rho_{AHE}^{sh} = 1.45\ n\Omega * cm$. For the planar Hall measurements, the Hall signal is recorded as a function of $\varphi$, the azimuthal angle of an in-plane magnetic field with constant strength larger than the perpendicular anisotropic field. As shown in Fig. 1(c), by fitting the planar Hall resistivity using a $sin(2\varphi)$ function, the planar Hall resistivity modulation is extracted to be $\Delta\rho_{PH} = 23.54\ n\Omega * cm$. We also measure the longitudinal MR as a function of an in-plane magnetic field oriented in different directions. During the field sweeping from 0 to ±1 T, the TIG magnetization rotates from out-of-plane to in-plane. As summarized in Fig. 1(d), when the field is in the y-direction ($\varphi = 90^o$), the MR is much larger than when the field is in the x-direction ($\varphi = 0^o$). This indicates the dominance of the SMR over the anisotropic magnetoresistance; therefore, the planar Hall resistivity $\Delta\rho_{PH}$ should just be approximately equal to $\Delta\rho_{SMR}$. Indeed, we find that $\Delta\rho_{SMR}/\rho \approx \Delta\rho_{PH}/\rho = 5.0 \times 10^{-4}$. When the field is applied at $\varphi = 45^o$ and $135^o$, the MR is about an half of that in the x- or y-axis, which is expected from the $m_y^2$-dependence. It is important to note that $\Delta\rho_{SMR}$ is 16.2 times larger than $\Delta\rho_{AHE}^{sh}$.

Next we turn to the experimental determination of the effective fields of DLT and FLT using harmonic Hall measurements. This method has been widely used to extract the value of $H_{FL}$ and $H_{DL}$ in FM metal/HM bilayers [7, 28-30]. However, this field-sweep method is not suitable here because the 2ω Hall signals under an x-axis field also contain a strong spin-dependent thermal signal due to Joule heating [23, 31-33], which makes it very difficult to accurately determine effective field of FLT. Here, we extract $H_{FL}$ and $H_{DL}$ of SOTs by measuring the angle-dependent harmonic Hall responses instead [34, 35] and separate various contributions from DLT, FLT and the spin-dependent thermal effect. The 2ω Hall resistance signal $R_H^{2w}$ contains both $cos\varphi$ and $cos2\varphi cos\varphi$ terms,

$$R_H^{2w} = (R_{DL}^{2w} + R_{TH}^{2w}) \cdot cos\varphi + (R_{FL}^{2w} + R_{Oe}^{2w}) \cdot cos2\varphi \cdot cos\varphi, \quad (2)$$

where $R_{DL}^{2w}$ and $R_{TH}^{2w}$ are the 2ω Hall resistances from the DLT and thermal effect contributions, respectively, while $R_{FL}^{2w}$ and $R_{Oe}^{2w}$ correspond to the FLT and Oersted field contributions, respectively. As shown in Fig. 2(b), the angular dependence of the $R_H^{2w}$ is first measured under different constant magnetic field strengths. $(R_{DL}^{2w} + R_{TH}^{2w})$ and $(R_{FL}^{2w} + R_{Oe}^{2w})$ are then extracted by fitting $R_H^{2w}$ using Eq. (2). After that, $R_{DL}^{2w}$ and $R_{TH}^{2w}$ are further separated from each other according to their respective field-dependent behaviors as described in SM [23]. Fig. 3(a) shows $(R_{DL}^{2w} + R_{TH}^{2w})$ as a linear function of $1/\mu_0(H + H_k)$ with intercept $R_{TH}^{2w}$ and slope proportional to $H_{DL}$. Here $\mu_0$ is the vacuum permeability, $H_k$ is the anisotropic field determined from independent spin-torque ferromagnetic resonance measurements [23, 36, 37]. Similarly, in Fig. 3 (b), $(R_{FL}^{2w} + R_{Oe}^{2w})$ depends linearly on $1/\mu_0 H$ with the slope containing $H_{FL} + H_{Oe}$. After subtracting $H_{Oe}$ (see SM [23]), we obtain $H_{FL}$ for different currents. Fig. 3(c) summarizes both effective fields of SOTs as a function of the current density $J$. Clearly, both $H_{DL}$ and $H_{FL}$ depend linearly on $J$ for small current densities. From the slopes we extract the effective fields of DLT and FLT per unit current density: $\partial_J H_{DL} = 117$ Oe/$(10^{11} A/m^2)$ and $\partial_J H_{FL} = 20$ Oe/$(10^{11} A/m^2)$, for TIG(4.8 nm)/Pt(4 nm).

With the extracted SH-AHE, SMR, and effective fields of DLT and FLT in TIG(4.8 nm)/Pt(4 nm), we can test the validity of Eq. (1). We find that $\partial_J H_{FL}/\partial_J H_{DL} = 0.17$, which is 2.7 times larger than $\Delta\rho_{AHE}^{sh}/\Delta\rho_{SMR} = 0.062$. This significant discrepancy is well beyond the experimental uncertainty. To further confirm this point, we perform the same measurements in all TIG($t_{TIG}$ nm)/Pt(4 nm) heterostructures with $t_{TIG}$ ranging from 3.2 to 9.6 nm and summarize the $t_{TIG}$-dependence of $\partial_J H_{SOT}^{eff}$ in Fig. 4(a). First of all, the magnitude of both $\partial_J H_{DL}$ and $\partial_J H_{FL}$ decreases quickly with increasing $t_{TIG}$ and approaches saturation as $t_{TIG} > 6.4$ nm. Furthermore, similar to $t_{TIG}$=4.8 nm sample, $\partial_J H_{DL}$ is much larger than $\partial_J H_{FL}$ for three other samples. The inset of Fig. 4(a) shows both SH-AHE and SMR vs. $t_{TIG}$ for all samples, $\Delta\rho_{SMR}/\rho$ is much larger than $\Delta\rho_{AHE}^{sh}/\rho$ across the $t_{TIG}$-thickness range. Quantitatively, Fig. 4(b) displays both ratios of $\partial_J H_{FL}/\partial_J H_{DL}$ and $\Delta\rho_{AHE}^{sh}/\Delta\rho_{SMR}$ vs. $t_{TIG}$ for all four samples. Most importantly, the $\partial_J H_{FL}/\partial_J H_{DL}$ curve stays above that of $\Delta\rho_{AHE}^{sh}/\Delta\rho_{SMR}$ and the actual values differ by at least a factor of two with each other. This severe discrepancy further confirms the breakdown of Eq. (1).

Apart from comparing the two ratios, we also examine the spin torque efficiency for both DLT and FLT. On one hand, the spin torque efficiency of DLT and FLT can be determined from the harmonic Hall measurements, i.e. $\xi_{DL(FL)} = \frac{2e}{\hbar}\mu_0 M_s t_{FM}^{eff} \frac{H_{DL(FL)}}{J}$ [30, 38-40], where $t_{FM}^{eff}$ and $M_s$ are the effective thickness and saturation magnetization of the FM layer. These two efficiencies are found to be $\xi_{DL} = 0.058$ and $\xi_{FL} = 0.0077$ for TIG(3.2 nm)/Pt(4 nm) heterostructure. On the other hand, they can also be estimated from measured SMR and SH-AHE: $\xi_{DL} = \frac{\Delta\rho_{SMR}/\rho}{\theta_{SH}\frac{\lambda}{t_{HM}}\tanh\left(\frac{t_{HM}}{2\lambda}\right)}$ and $\xi_{FL} = \frac{\Delta\rho_{AHE}^{sh}/\rho}{\theta_{SH}\frac{\lambda}{t_{HM}}\tanh\left(\frac{t_{HM}}{2\lambda}\right)}$ (see SM [23]), where $\rho$, $\lambda$ and $t_{HM}$ are resistivity, spin diffusion length and HM thickness, respectively. As discussed in SM [23], if we take relatively small values of $\theta_{SH} = 0.06$ and $\lambda = 0.6\ nm$ which gives the upper bounds for $\xi_{DL}(\xi_{FL})$ which are estimated to be: 0.0596(0.0037) for TIG(3.2 nm)/Pt(4 nm). It is important to note that the two $\xi_{DL}$ values obtained from the two different measurements agree with each other, which seemly conforms the validity of the model, but the maximum $\xi_{FL}$ estimated from the SH-AHE is only less than a half of that from the FLT effective field. Other choices of $\theta_{SH}$ and $\lambda$ would lead to smaller $\xi_{DL}$ and $\xi_{FL}$, which means an even larger discrepancy for the FLT efficiency. A similar conclusion can be drawn for other samples since the observed nearly linear $H_{DL(FL)}/J$ vs. $1/M_s t_{TIG}$ behavior (see SM [23]) implies $t_{TIG}$-independent $\xi_{DL}$ and $\xi_{FL}$. Therefore, we conclude that the bulk SHE model significantly underestimates FLT efficiency for all samples.

These two outstanding discrepancies highlight the deficiency of the bulk SHE model for angular momentum transfer in MI/HM heterostructures. In particular, the bulk SHE alone fails to account for the observed large magnitude of the FLT efficiency, which suggests that this picture neglects some essential ingredients. Since the bulk SHE model only considers the spin current generation, diffusion, and drift in the HM layer, the MI/HM interfacial effects are not included [41]. The Rashba effect from broken inversion symmetry at interfaces is the first candidate. In fact, this effect was first proposed for asymmetric film structures and shortly observed in Pt/CoFeB/AlO$_x$ [42]. An in-plane effective Rashba field as large as 3-10 kOe was estimated from the experimental data [4]. The presence of this effect can in principle significantly contribute to the effective FLT efficiency. A second possibility is the magnetic proximity effect at the MI/HM interface. It is known that Pt is prone to be magnetized by proximity coupling. Just as in the metallic FM/HM systems that have larger FLT [42], the induced ferromagnetism at the interface resembles a metallic FM/HM bilayer inserted in the MI/HM heterostructure and therefore likely adopts higher FLT efficiency of the FM/HM system. Above all, both interfacial effects can possibly lead to a significantly stronger FLT as well as breakdown of Eq. (1).

In summary, we have quantitatively examined the validity of the bulk SHE model for SMR and SOTs in TIG/Pt heterostructures and found that the ratio between effective fields of FLT ($H_{FL}$) and DLT ($H_{DL}$) is at least two times larger than the ratio between SH-AHE and SMR. In addition, the model greatly underestimates FLT efficiency relative to DLT. These significant discrepancies suggest that interfacial mechanisms such as the Rashba effect and magnetic proximity effect play a more important role in generating larger FLT.

We would like to thanks Gen Yin and Yabin Fan for useful discussions. The work was supported as part of the SHINES, an Energy Frontier Research Center funded by the U.S. Department of Energy, Office of Science, Basic Energy Sciences under Award No. SC0012670.

harmonic Hall measurements and magneto-transport measurements, and estimation of spin-torque efficiencies from both SOTs, SMR and SH-AHE.

[37] C. He, A. Navabi, Q. M. Shao, G. Yu, D. Wu, W. Zhu, C. Zheng, X. Li, Q. L. He, S. A. Razavi, K. L. Wong, Z. Zhang, P. K. Amiri, and K. L. Wang, Appl. Phys. Lett. **109**, 202404 (2016).

[38] C.-F. Pai, Y. Ou, L. H. Vilela-Leão, D. C. Ralph, and R. A. Buhrman, Phys. Rev. B **92**, 064426 (2015).

[39] M.-H. Nguyen, D. C. Ralph, and R. A. Buhrman, Phys. Rev. Lett. **116**, 126601 (2016).

[40] W. Zhang, W. Han, X. Jiang, S.-H. Yang, and S. S. P. Parkin, Nat. Phys. **11**, 496 (2015).

[41] X. Fan, H. Celik, J. Wu, C. Ni, K.-J. Lee, V. O. Lorenz, and J. Q. Xiao, Nat. Commun. **5**, 3042 (2014).

[42] I. M. Miron, Gilles Gaudin, S. Auffret, B. Rodmacq, A. Schuhl, S. Pizzini, J. Vogel, and P. Gambardella, Nature Mater. **9**, 230 (2010).11

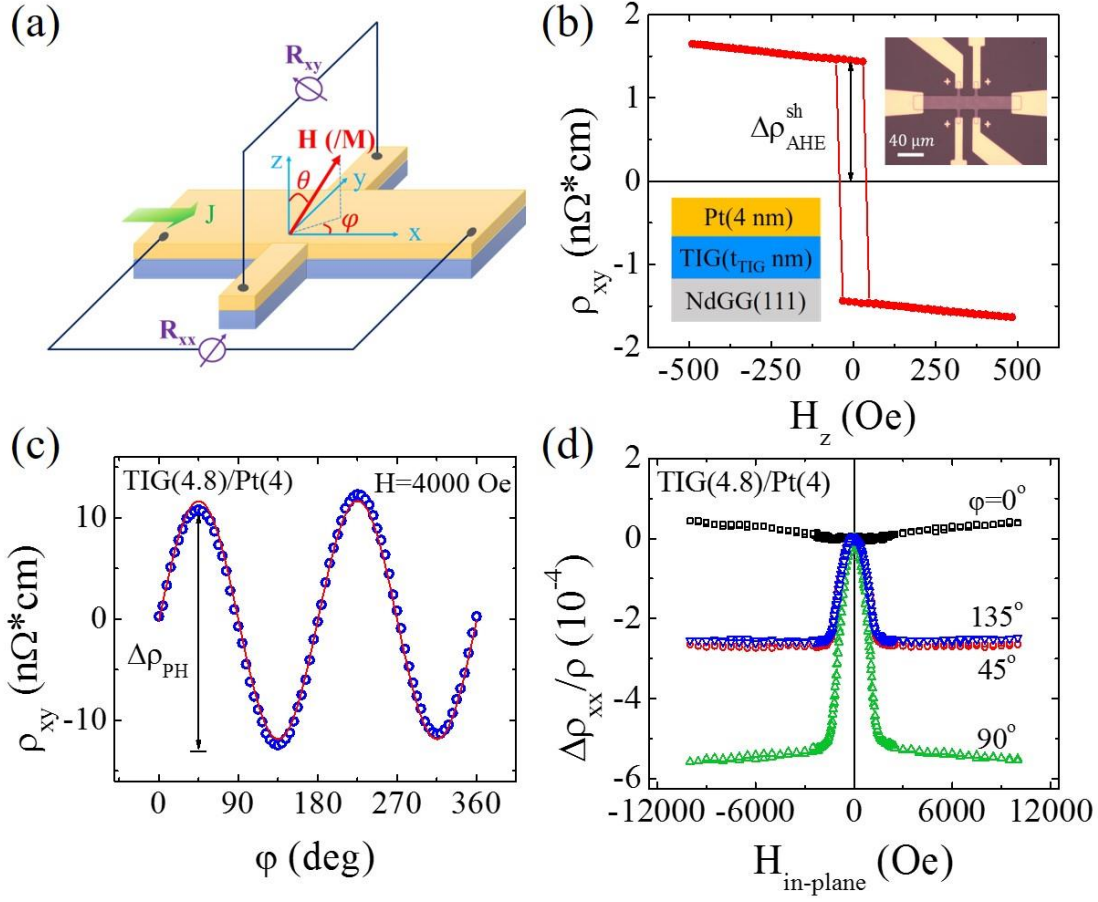

FIG. 1. SH-AHE and magnetoresistance of TIG(4.8 nm)/Pt(4 nm) heterostructure. (a) Schematic drawing of the Hall device and measurement geometry. (b) AHE hysteresis loop marked with its magnitude $\Delta\rho_{AHE}^{sh}$. The insets show stack structure and optical image of the Hall bar. (c) Angular dependence of planar Hall resistivity with a 4000 Oe in-plane rotating magnetic field. The red solid line is a fit with $\sin(2\varphi)$. The arrows indicate the resistivity modulation due to the planar Hall effect, $\Delta\rho_{PH}$. (d) longitudinal magnetoresistance as a function of in-plane magnetic field with orientations of $\varphi = 0^o$, $45^o$, $90^o$ and $135^o$.

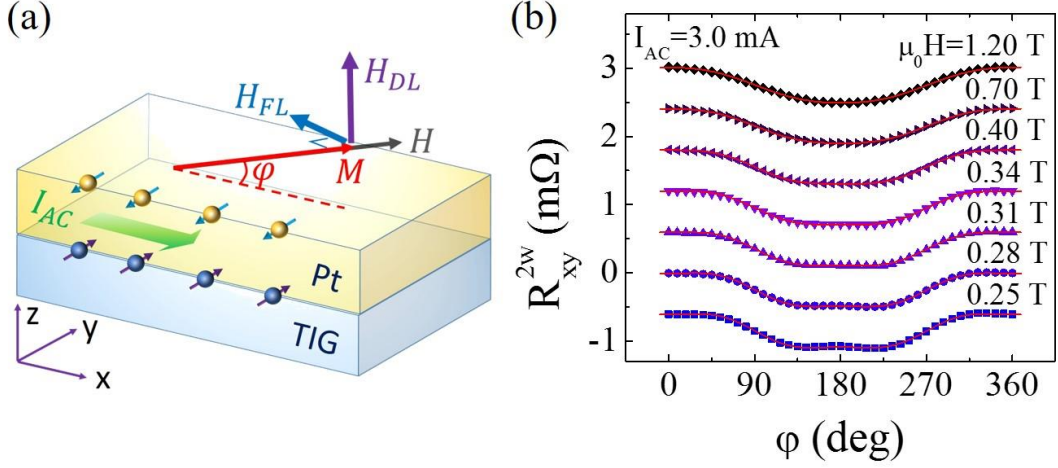

FIG. 2. (a) Schematic illustration of two effective fields $H_{DL}$ and $H_{FL}$ in harmonic Hall measurement. $I_{Ac}$ is the amplitude of AC current. H and M are the in-plane magnetic field and magnetization, $\varphi$ is the azimuthal angle. (b) Representative angle-dependent 2ω-Hall signals under different in-plane magnetic fields for TIG(4.8 nm)/Pt(4 nm). The rms amplitude of the AC current is 3.0 mA, corresponding to a current density of $J = 0.375 \times 10^{11}\ A/m^2$. For clarity, the curves are vertically shifted. The red lines are fits using Eq. (2).

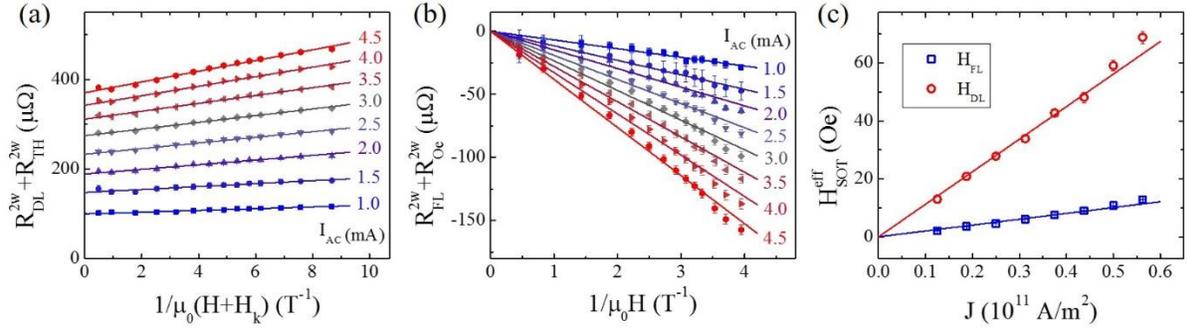

FIG. 3. (a) 2ω Hall resistance from DLT and spin-dependent thermal effect as a function of $[\mu_0(H+H_k)]^{-1}$, where $H_k$ is the effective anisotropic field including the demagnetizing field and the perpendicular magnetic anisotropy field. (b) 2ω Hall resistance from of FLT and the Oersted field as a function of $(\mu_0 H)^{-1}$. (c) Current density dependence of the effective fields of SOTs. The solid lines in (a), (b) and (c) are linear fits.

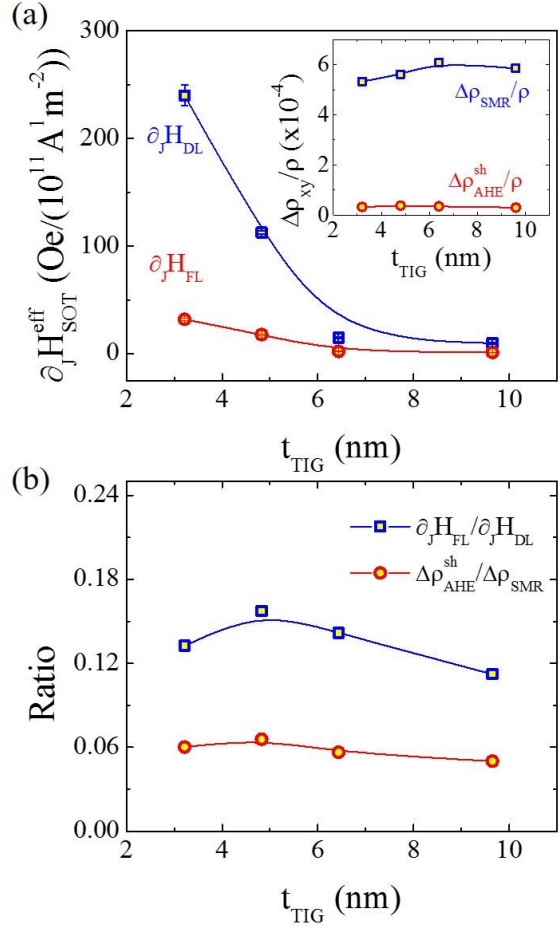

FIG. 4. (a) TIG thickness dependence of effective fields of SOTs per unit current density with the inset showing both SMR and SH-AHE as a function of TIG thickness. (b) $\partial_J H_{FL}/\partial_J H_{DL}$ and $\Delta\rho_{AHE}^{sh}/\Delta\rho_{SMR}$ as a function of TIG thickness. The solid lines are guides to the eye.